# How Embeddedness Affects the Evolution of Collaboration: The Role of Knowledge Stock and Social Interactions

**Hongshu, Chen**     Beijing Institute of Technology, People's Republic of China | Hongshu.chen@bit.edu.cn

**Qianqian, Jin**     Beijing Institute of Technology, People's Republic of China | qianqian_jin@ bit.edu.cn

**Xuefeng, Wang**     Beijing Institute of Technology, People's Republic of China | wxf5122@bit.edu.cn

## ABSTRACT

Science and technology are becoming increasingly collaborative. This paper aims to explore the factors and mechanisms that impact the dynamic changes of collaborative innovation networks. We consider both collaborative interactions of organizations and their knowledge element exchanges to reveal how social and knowledge network embeddedness affects the collaboration dynamics. Knowledge elements are extracted to present the core concepts of scientific and technical information, overcoming the limitations of using predefined categorizations such as IPC when representing the content. Based on multiple collaboration and knowledge networks, we then conduct a longitudinal analysis and apply a stochastic actor-oriented model (SAOM) to model network dynamics over different periods. The influence of network features and structures, individual node characteristics, and various dimensions of proximity on collaboration dynamics is tested and analyzed.

## INTRODUCTION
With the increasing refinement and complexity of scientific and technological issues, the innovation landscape has undergone a shift from individual-led research to collaborative innovation (Ba et al., 2021). Through collaboration, knowledge of organizations can be socially validated and integrated with that of others to recombine existing knowledge and create new cognitions (Massey et al., 2006; Nonaka et al., 2009). Therefore, innovative organizations are embedded within multiple heterogeneous networks, which include not only collaboration networks that reflect social interactions but also knowledge networks formed when exchanging knowledge elements during collaborative innovation (Brennecke et al., 2017; Liu et al., 2022).

In the process, collaboration relationships may solidify, transform, or collapse over time (Park et al., 2020), sparking widespread interest in measuring the factors that affect innovative collaboration. Existing research has explored how collaborations form and evolve by modeling the pure social interaction of organizations using the co-applications of patents, co-authorship of research articles or co-funded grants, with promising results (Akhtar et al., 2019; Huang et al., 2015; Yan et al., 2018). However, the knowledge stock of organizations and the exchange of knowledge elements are rarely taken into consideration when revealing the driving forces of collaboration. To do so, extracting knowledge elements with an appropriate granularity to reflect knowledge interactions also warrants further research.

## METHODOLOGY
Facing these challenges, we consider both collaborative interactions between organizations and their knowledge element exchanges to reveal how social and knowledge network embeddedness affects the collaboration dynamics. This methodology framework is shown in Figure 1. The data used in this research are patent data drawn from the Derwent Innovation Index database (DII). To chart the evolution of collaborations, we construct multiple subsets of patent data for different time periods. For dataset in each period, multiple networks are constructed, including a collaborative innovation network of organizations, a global knowledge network of all knowledge elements in the corresponding patent dataset, and local knowledge networks for each organization. We then perform a longitudinal analysis and apply a stochastic actor-oriented model (SAOM) to model network dynamics over different periods.

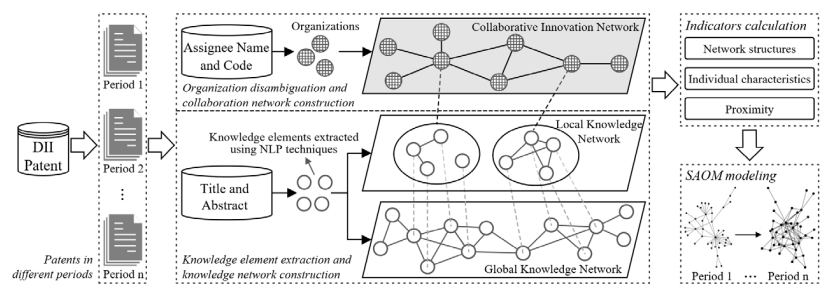

**Figure 1. Methodology framework: multi-embeddedness of collaborative and knowledge networks**

Knowledge element (KEs) extraction is the foundation for reflecting knowledge stock and exchanges. The prior literature considers IPC codes (Wang et al., 2014), keywords(Guan et al., 2017), topics or predefined tags (Li et al., 2023) as a proxy for knowledge elements. Yet both predefined taxonomies or rough-grained document comprehension have limitations in directly indicating fundamental content. We refine the granularity of knowledge elements to $n$-gram terms in this paper ($n = 2,3,4$). First, we use GPT-3.5 to extract noun phrases, and perform lemmatization and stemming to reduce words to their base form. Phrases that have similar meanings are merged and consolidated. The TF-IDF values are computed to keep the elements with higher statistical significance. The SAOM is then applied to test the factors that drive network dynamics (Snijders, 2017). Our assumptions revolve around the influence of features and structures, individual node characteristics, and various dimensions of proximity on collaboration dynamics (Arts et al., 2021; Chen et al., 2022; Guo et al., 2021; Heirman et al., 2015; Tsouri et al., 2022; Xiao et al., 2022; Zhang et al., 2022), as shown in Table 1.

| Effect | Indicator | Description |
|---|---|---|
| Collaboration network | Node degree | The number of organizations' partners |
| | Transitive triads | Network closure |
| Global knowledge network | Combinatorial potential | Average degree centralities of KEs |
| | Combinatorial opportunity | Average structural holes of KEs |
| Local knowledge network | Modularity | The presence of communities or modules |
| | Global clustering coefficient | The number of closed triplets divided by sums of triplets |
| Individual characteristics | Knowledge diversity | Determined by both the number and semantics of KEs |
| | Knowledge uniqueness | Determined by the frequency of KEs |
| | Organization Type | Firms or academic institutions |
| Proximity | Cognitive proximity | Cosine similarity of semantic vectors organizations' KEs |
| | Organizational proximity | Whether two organizations are in the same type |

Table 1. The variables of the SAOM

**EMPIRICAL STUDY AND RESULT**

We retrieve 28,905 patents granted from 2003 to 2022 in the field of lithography, which is a crucial area of semiconductor manufacturing, for empirical study. Further filtering is applied to select assignees that had participated in at least one collaboration and had a minimum of 2 granted patents. This resulted in 346 assignees and 22,448 patents. During knowledge elements extraction, elements with a TF-IDF value below 0.001 are removed. There are 7,022 knowledge elements were kept to construct local and global networks. The result shows that both global and local knowledge network embeddedness of the knowledge elements for an assignee have impact on collaboration dynamics. Organizations that possess knowledge with higher combinatorial potential are more likely to collaborate. If an assignee has a greater number of unique and less diverse knowledge elements, or if its local knowledge network exhibits stronger connectivity, its willingness to add a new tie or disconnect existing ones decreases. In this field, compared to academic organizations, enterprises are less willing to collaborate. Both cognitive and institutional proximity drive partnership building.

| Variable | Estimate | SE | Variable | Estimate | SE |
|---|---|---|---|---|---|
| Rate period 1 | 3.0534 | 0.404 | Modularity | 0.3957 | 1.125 |
| Rate period 2 | 2.8015 | 0.438 | Global clustering coefficient | -4.0214*** | 1.506 |
| Rate period 3 | 1.3126 | 0.193 | Knowledge diversity | 0.0149* | 0.009 |
| Degree (density effect) | -5.8416*** | 0.158 | Knowledge uniqueness | -2.55* | 1.366 |
| Transitive triads | 1.9785*** | 0.153 | Organization type | -1.7902*** | 0.449 |
| Combinatorial potential | 10.952* | 6.121 | Organizational proximity | 0.3099** | 0.154 |
| Combinatorial opportunity | 10.7557 | 16.425 | Cognitive proximity | 2.7091*** | 0.553 |

Note: *** $p<0.01$, ** $p<0.05$, * $p<0.1$. SE means Standard Error. Overall maximum convergence ratio is 0.1068.
Parameter setting: seed = 8, n3 = 3000, nsub = 5

Table 2. Results of the SAOM

**CONCLUSION**

This paper explores both collaborative interactions of organizations and their knowledge element exchanges to reveal how social and knowledge factors affect the dynamics of collaboration. We first contribute to extracting knowledge elements to present core concepts of patents, overcoming the limitations of using the IPC when representing the content. We then contribute to modeling collaboration dynamics by performing a longitudinal analysis on multiple collaboration and knowledge networks over different periods, in which network structures, node characteristics, and different dimensions of proximity on collaboration dynamics are tested and analyzed.